\documentclass[showpacs,preprintnumbers,amsmath,
amssymb]{revtex4}
\usepackage{graphicx}
\usepackage{dcolumn}
\usepackage{bm}
\usepackage{tensor}

\begin{document}

\title{Coherency Tuning via Phase Conjugation \\
in Fiber Laser Networks.}

\author{A.Yu.Okulov}
\email{alexey.okulov@gmail.com}
\homepage{http://okulov-official.narod.ru }
\affiliation{Russian Academy of Sciences, 119991,  Moscow, Russian Federation.}

\date{\ December 20, 2014}
 
\begin{abstract}
{Phase-locking of large mode area 
Yb photonic crystal fiber amplifiers 
via stimulated Brillouin scattering phase conjugation 
is analyzed. The wavefront reversing mirrors are proposed 
for chirped pulse fiber laser networks. The chirp control 
via both nonresonant Kerr and resonant index is considered. 
}

\end{abstract}

\maketitle

\vspace{1cm}

\maketitle 
The networks of fiber lasers with kilohertz repetition rates 
promise a revolutionary characteristics of particle accelerators 
and high energy density facilities \cite {Mourou:2013,Bulanov:2006}. 
For this purpose 
a lattice of the fiber amplifiers emitting a thousands of beamlets 
of $mJ$ pulses with identical chirped spectrum could maintain a high spatial 
and temporal coherency under accurate ($\lambda/10$ - $\lambda/20$) 
tuning of optical paths \cite {Brignon:2011}. 
   
The equalization of optical paths difference $\Delta L$ 
in laser sets might be achieved 
by means of phase conjugation ($PC$) 
as well \cite {Fisher:1983,Brignon:2013}. 
This wavefront reserving technique when backwardly amplified 
wave $\bf E_b$ propagates as $time$ $reversal$ replica of incident 
wave $\bf E_f$ provides a unique opportunity for coherency 
protection in fiber amplifier sets \cite {Okulov:2014}. When 
phase-conjugated mirrors with Stimulated Brillouin scattering are 
used the phase lag $\Delta \phi=\Delta k \Delta L$ due to 
gigahertz frequency difference between incident and $PC$ reflected   
wave $\Delta k= k_f - k_b $ may be effectively used for decoupling 
of master oscillator and amplifier set \cite {Basov:1980} 
when $\Delta \phi=\pi$. In this 
robust $PC$ interferometry the amplifier set 
is automatically self-adjusted even for the low quality 
optical components \cite {Rockwell:1986}. 
  
In addition to recent proposal of chirped pulse $PC$ mirrors with 
instantaneous $Kerr$ response \cite {Okulov:2014} via 
degenerate four-wave mixing ($DFWM$) one may suggest 
the fiber amplifier set (fig.1) with  
Brillouin enhanced four-wave mixing $PCM$ \cite {Boyd:1998} in order to 
improve the $PC$ reflectivity and stability with respect variations of 
$DFWM$ $PCM$ beams. The duration of master oscillator 
is taken to avoid random phase jumps uniformly distributed 
within interval $[-\pi,\pi]$  of $PC$ reflected $SBS$ 
wave  \cite {Basov_Lett:1980} .
In this scheme 
the transform limited (i.e. nonchirped) envelope laser pulse of nanosecond 
duration $\tau \sim 1$ $nsec$ is injected into binary tree Michelson 
amplifier set and backwardly reflected from Brillouin $PCM$ \cite {Okulov_2:1980}.
Because $\tau$ is significantly smaller than acoustical phonon lifetime 
$\tau_{ph}$ \cite {Okulov:1983} the acoustical grating in $PCM$ 
is $frozen$ during reflection  \cite {Boyd:2007}. Thus temporal envelope 
of reflected wave will be identical to the 
incident one:
\begin{eqnarray}
\label{envelope_gauss_secant}
{E_{PC}}(z,t,{\vec r})= f(z \pm ct){\bf E_f}^{*}(z,{\vec r}) 
&& \nonumber \\
\cong  f(z \pm ct) \sum_{m,n} {a^{*}_{m,n}}(z,{\vec r})
\cdot \exp {\:}[-i \Delta \Phi_{m,n}] , {\:} {\:} {\:} 
\end{eqnarray}
where $f(z \pm ct)$ is $transform$ $limited$ 
temporal envelope of the master oscillator pulse,
 ${\bf E_f}(z,{\vec r})={\sum a_{m,n}}(z,{\vec r})
\cdot \exp {\:}[i \Delta \Phi_{m,n}]$ is spatial slowly 
varying envelope, $\Delta \Phi_{m,n}$ are phase piston errors 
compensated due to $SBS$ $PCM$. The phase piston errors are 
responsible for decoherence of laser arrays 
\cite {Okulov:1991} and filamention of the single lobe far field 
pattern \cite {Okulov:1993}. 

The mechanism of high fidelity of $PCM$ reflection is in propagation 
of Stockes 
component inside randomly spaced helical waveguides 
forming speckle pattern \cite {Okulov:2009}. The helicity of SBS 
speckle phase conjugated 
pattern \cite {Woerdemann:2009} 
is due to alternation of the orbital angular of 
light in phase conjugation \cite {Okulov:2008,Okulov:2012josa} and 
excitation of acoustical vortices carrying orbital angular 
momentum \cite {Okulov:2008J}. 

In order to realize the basic advantages of chirped 
pulse compression technique \cite {Strickland_Mourou:1985} 
the chirping frequency modulation 
might be produced via 
combined action of $\chi^3$ nonresonant 
Kerr index $f(z + ct) \exp [i k n_2 \ln |f(z+ct)|^2]$ 
and modulation via resonant index of $Yb$ ions changes 
due to backward propagation in saturated fiber 
amplifier (fig.1). The chirped self phase 
modulation of backwardly amplified wave $\bf E_b$ 
is described by nonlinear Shrodinger-Frantz-Nodvik 
equation \cite {Okulov:1988_QE}:
\begin{eqnarray}
\label{nanosec}
\ {\frac {\partial {{\bf E_b}(z,t,\vec r )}} {\partial z} } - 
{\frac {n_r} {c} }{\frac {\partial {{\bf E_b}}} {\partial t} }+
{\frac {i}{2 k_b}} {\nabla}_{\bot}^{{\:}2} {\bf E_b} =
&& \nonumber \\
{\frac { \sigma_{_{Yb}}{\:}{N{(z,t)}}} {2{\:}  } } 
{\bf E_b}(1+i \Delta \omega T_2)+ i k n_2 |{\bf E_b}|^2 \cdot {\bf E_b}, 
\end{eqnarray}
with  dynamics of population 
inversion $N(z,t)$ given by:
\begin{equation}
\label{inversion nanosec}
\ {\frac {\partial N{(z,t)}} {\partial t}} =
- \sigma_{_{Yb}} N(z,t) \cdot |{\bf E_b}|^2  +
{\frac { {N_0{(z)}-N{(z,t)}}} {T_1} },
\end{equation}
where $\sigma_{_{Yb}} \sim 10^{-20} cm^2$ is stimulated 
cross section of $Yb^{3+}$ resonant transition, $T_1, T_2$ are 
longitudinal and transversal relaxation times 
of $Yb$ ions,
$\Delta \omega$ is detuning of carrier frequency of 
$\bf E_b$  from the center of gain line 
of $Yb$ ions.

The exact solution taking into account 
gain saturation of the nonlinear Shrodinger-Frantz-Nodvic 
may be obtained in explicit form 
because for $\tau \sim 10^{-9}$ $sec$ and for the typical distance 
between of fiber amplifiers set and PCM $> 15 cm$ the backward 
wave ${\bf E_b}$ is amplified without interference with 
forward wave ${\bf E_f}$ \cite {Okulov:2014}: 
\begin{widetext}
\begin{equation}
\label{pulse_chirping}
{\bf E_b}(z,t )={\frac {{\bf E_b}
(z=0,t )
\cdot exp{\:}[-i k n_2 \int_{_{_{0}}}^{{\:}L_f} 
[|{\bf E_b}(\eta ,t )|^2  - 
i \Delta \omega T_2  \sigma T_1 |{\bf E_b}(\eta ,t )|^2]d \eta ]}
{\sqrt{1-[1- exp{\:}[-\sigma_{_{Yb}} 
\int_{0}^{L_{f}} N_0(z^{'})d z^{'}]
{\:} exp{\:}[{\:}-2\sigma_{_{Yb}} 
\int_{-\infty}^{t-z/v_g}{{|\bf E_b}(z,\tau)|}^2 
d \tau]] } }},
\end{equation}
\end{widetext}

where $L_f$ is average length of fiber amplifier. 
The solution (\ref {pulse_chirping}) was obtained 
using procedure described in \cite {Basov:1966} is valid for short 
($L_f \sim 2 m$) large mode area $mJ$-level 
fiber amplifier \cite{Eidam:2011}. 
 The figure of merit ($FOM$) of 
phase-locked single spatial mode output is 
evaluated \cite{Okulov:2014}.

\begin{figure}
\center{\includegraphics[width=14cm]{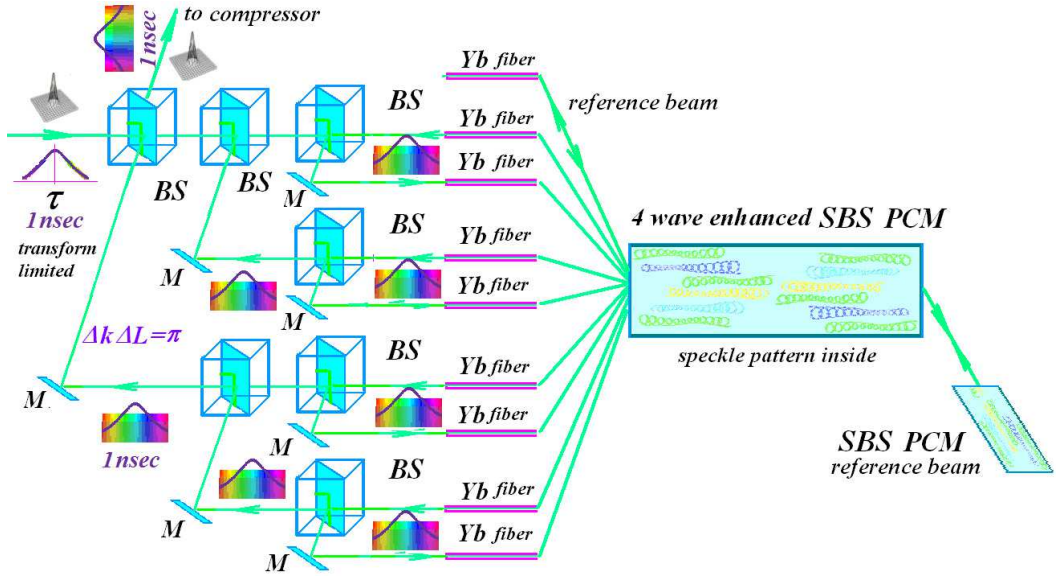}}
\caption{Michelson binary tree for coherent summation of $\bf Yb$ fiber 
amplifiers lattice. $\bf BS$ are 50/50 beamsplitters, M are mirrors, 
$\Delta k \Delta L=\pi$  is output decoupling 
phase lag due to $\bf SBS$ $\bf PCM$ 
 reflection \cite {Basov:1980}, $\bf SBS$ $\bf PCM$ is 
broad area lightguide SBS four wave mixing enhanced 
mirror \cite {Boyd:1998} with speckle pattern 
produced by $\bf Yb$ fiber amplifiers set , 
the spectrum colors with 
imposed envelope show the $\bf PCM$ reflected pulses whose chirp is 
induced by backward propagation in 
resonant gain medium \cite {Okulov:1988_QE}. 
$\bf SBS$ $\bf PCM$ $\bf reference$ 
is mirror for reference waves $E_{ref_f}$ $E_{ref_b}$.}
\label{fig.1} 
\end{figure}
\end{document}